\documentclass[12pt,english]{article}
\usepackage{babel}
\usepackage{amsmath}
\usepackage{amssymb} 
\usepackage[small]{caption2} 
\usepackage{fleqn} 
\usepackage{graphicx} 
\usepackage[small,loose]{subfigure}  
\usepackage{cite} 
\advance \headheight by 3.0truept       
\setcaptionwidth{.85\textwidth}

\newcommand{\CenterObject}[1]{\ensuremath{\vcenter{\hbox{#1}}}}

\makeatother

\begin{document}
\title{
\begin{flushleft}
{\normalsize DESY 04-216\hfill{November 2004}}
\end{flushleft}
\vspace{2cm}
{\bf Maximal Temperature \\in Flux Compactifications}\\[0.8cm]}
\author{\\
{\bf\normalsize 
Wilfried Buchm\"uller, Koichi Hamaguchi, Oleg Lebedev} \\[0.2cm] 
{\it\normalsize Deutsches Elektronen-Synchrotron DESY, 22603 Hamburg, Germany}
\\[0.4cm] 
{\bf\normalsize Michael Ratz}\\[0.2cm]
{\it\normalsize Physikalisches Institut der Universit\"at Bonn,}\\
{\it\normalsize Nussallee 12, 53115 Bonn, Germany}
 }
\date{}
\maketitle \thispagestyle{empty} 

\abstract{
Thermal corrections have an important effect on moduli stabilization
leading to the existence of a maximal temperature, beyond which
the compact dimensions decompactify.
In this note, we discuss generality  of our earlier analysis 
and apply it to the case of flux compactifications.
The maximal temperature is again found to be 
controlled by the supersymmetry breaking scale,
$T_{\rm crit} \sim \sqrt{m_{3/2}M_\mathrm{P}}$.} 

\clearpage

In string theory gauge and Yukawa couplings are dynamical quantities whose
values are determined by expectation values of scalar fields (moduli).
In perturbation theory their potential is flat, and their stabilization is
a topic of central importance in string theory. Recently, it has been 
proposed that a complete stabilization of all moduli fields 
\cite{Kachru:2003aw} is possible using a combination of fluxes  
\cite{Giddings:2001yu} and non-perturbative effects such as D--brane instantons
and gaugino condensation \cite{Derendinger:1985kk}.

The moduli potentials  receive important thermal corrections 
\cite{Buchmuller:2003is,Buchmuller:2004xr}. 
These corrections destabilize the moduli at sufficiently high temperature, 
i.e. drive them to the `run--away' minimum \cite{Dine:1985he}
where the compact dimensions decompactify. In order to avoid this
cosmological disaster the temperature in the early universe has to be
smaller than a maximal temperature $T_{\rm crit}$.

In the following we discuss generality of our previous analysis 
\cite{Buchmuller:2004xr} on dilaton destabilization by thermal effects 
and apply it to the KKLT scenario \cite{Kachru:2003aw}.
We find again that, generically, the maximal temperature is given by the 
scale of supersymmetry breaking, i.e., 
$T_{\rm crit} \sim \sqrt{m_{3/2}M_{\rm P}}$.
Temperature effects in some field theoretical flux compactifications
have recently been considered in \cite{Navarro:2004mm}.\\

\noindent
{\large\bf Moduli destabilization at high temperature}\\

In thermal field theory, the free energy $F$ plays the role of an effective 
potential. For a Yang-Mills theory at high temperature $T$, the free energy 
has a perturbative expansion in terms of the gauge coupling $g$,
\begin{equation}
F(g,T)\ =\ -{\pi^2 T^4 \over 24} \Bigl\{\alpha_0 + \alpha_2 g^2 + {\cal O}(g^3)
\Bigr\}\;.
\label{F}
\end{equation}
The crucial point of our subsequent discussion
is that $F$ increases with increasing $g$. For a
supersymmetric SU($N_c$) gauge theory with $N_f$ matter multiplets, 
one has (cf.~\cite{Kapusta:1989tk}) $\alpha_0=N_c^2 +2 N_c N_f -1$ and 
\begin{equation}
\alpha_2\ =\ -{3\over 8 \pi^2} (N_c^2-1)(N_c+3 N_f) < 0 \;.
\end{equation}
To order ${\cal O}(g^2)$ the free energy is determined by one- and two-loop 
diagrams (cf.~Fig.~1). The qualitative behaviour of the free energy,
$\partial F/\partial g^2 > 0$, is also valid at higher--loop level 
and even non-perturbatively for $g = {\cal O}(1)$ \cite{Buchmuller:2004xr}.
Furthermore, it holds for Abelian theories and for Yukawa couplings.  

\begin{figure}[!h]
 \centerline{
        \subfigure[Gauge-gauge.\label{fig:GaugeLoop}]{%
                \quad\quad
				\includegraphics{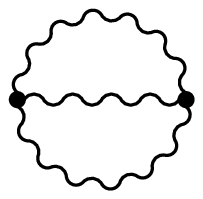}\hspace*{0.5cm}\includegraphics{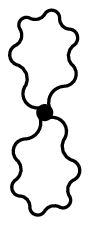}
				\quad\quad}
       \quad
        \subfigure[Gauge-matter.\label{fig:GaugeMatterLoop}]{%
                \includegraphics{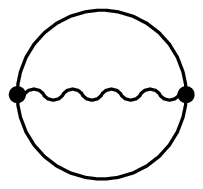}\hspace*{0.5cm}\includegraphics{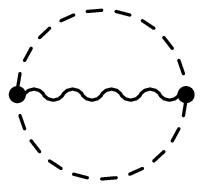}}
        \quad
        \subfigure[Matter-matter.\label{fig:MatterMatterLoop}]{%
                \quad\quad\includegraphics{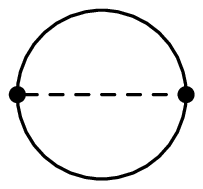}\quad\quad\quad}
 }
 \caption{Examples of two--loop diagrams contributing to the
effective potential.
 Wavy lines represent gauge fields, while matter fields are indicated by solid
 and dashed lines.}
 \label{fig:ThermLoops}
\end{figure}

In string theory the gauge coupling is related to some modulus $\Phi$,
\begin{equation}
g^2\ =\ {\kappa\over \Phi} \;,
\end{equation}
where $\kappa$ is a constant. Since thermal effects increase the effective 
potential, they will drive the modulus towards larger values, corresponding
to smaller couplings, and eventually destabilize the system. 
From Eq.~(\ref{F}) one obtains
\begin{equation}\label{vth}
 V_{\rm th}(\Phi,T) \equiv T^4  v_{\rm th}(\Phi)
 \ =\ 
 T^4\left(a_0 + a_2 {1\over \Phi} + \dots\right) \;,
\label{therm}
\end{equation}
where $a_0$ and $a_2$ are constants, with $a_2>0$. Clearly, the minimum of 
this potential is at $\Phi \rightarrow \infty$.

\begin{figure}[h]
 \centerline{\CenterObject{\includegraphics[scale=1]{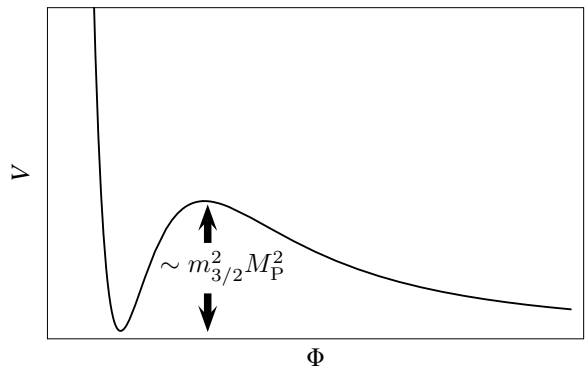}}}
 \caption{Typical supergravity potential for the modulus $\Phi$.
The local minimum is separated from the `run--away' one
by a barrier related to SUSY breaking.}
 \label{fig:fluxfig}
\end{figure}

The effective potential for $\Phi$ is the sum of the zero-temperature 
potential $V$ and the thermal correction,
\begin{equation}
 V_{\rm eff}(\Phi,T)
 \ =\ 
 V_{\rm th}(\Phi,T)+ V(\Phi) \;.
\end{equation}
The supergravity potential  $V(\Phi)$ for moduli is generated 
non--perturbatively. In known examples, it is related to supersymmetry
breaking at least for some moduli. Such potentials allow one to stabilize
the moduli at appropriate values, yet there is always the `run--away'
minimum at $\Phi \rightarrow \infty$, which is separated from the local 
minimum by a barrier related to SUSY breaking (Fig.~\ref{fig:fluxfig}). 
\vspace{5mm}

\begin{figure}[h]
 \begin{center}
 \subfigure[$V$.]{%
        \CenterObject{\includegraphics[scale=1]{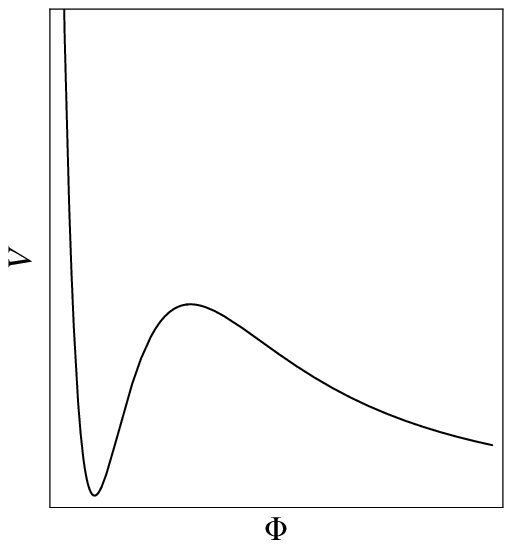}}}
 \hfil
 \subfigure[$V_\mathrm{th}$.]{%
        \CenterObject{\includegraphics[scale=1]{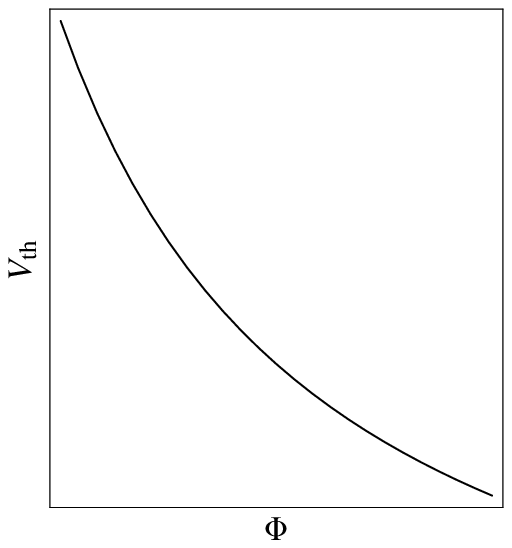}}}
 \hfil
 \subfigure[$V_\mathrm{eff}(\Phi,T)$.]{\CenterObject{%
        \includegraphics[scale=1]{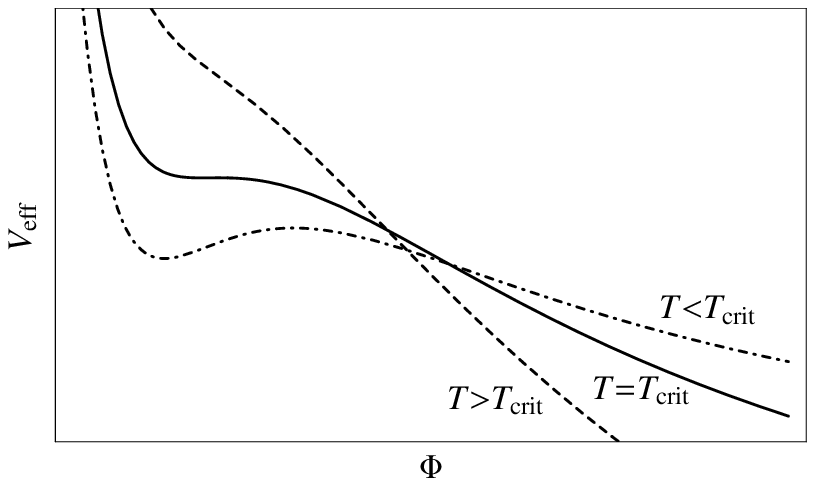}}}
\end{center}
 \caption{Moduli destabilization by temperature corrections. 
(a) supergravity potential,
(b) potential induced by thermal corrections,
(c) evolution of the full potential with increasing temperature: 
 $T<T_\mathrm{crit}$
 (dash-dotted curve), $T=T_\mathrm{crit}$ (solid curve), and
 $T>T_\mathrm{crit}$ (dashed curve).}
 \label{fig:ModuliDestablization}
\end{figure}

Clearly, when the temperature is high enough, the thermal potential 
(\ref{therm}) will dominate and drive $\Phi$ to infinity. Since the size of 
the potential near the local minimum is 
${\cal O}(m_{3/2}^2 M_{\rm P}^2)$, this occurs
at the critical temperature
\begin{equation}
 T_{\rm crit}
 \ \sim\ 
 \sqrt{m_{3/2} M_{\rm P}} \;,
\label{max}
\end{equation}
where $m_{3/2}$ is the gravitino mass and $M_{\rm P}=2.4\times 10^{18}$ GeV.
An example of moduli destabilization by thermal corrections is shown
in Fig.~\ref{fig:ModuliDestablization}.

The critical temperature is defined by the appearance of a saddle 
point at some value $\Phi_{\rm crit}$:
\begin{eqnarray}
 V'_{\rm eff}(\Phi_{\rm crit}, T_{\rm crit}) & =& 0 \;, \nonumber\\
 V''_{\rm eff}(\Phi_{\rm crit}, T_{\rm crit}) &= & 0 \;,
\end{eqnarray}
where the prime indicates differentiation with respect to $\Phi$.
Usually $V(\Phi)$ is a steep function, with exponential field
dependence, while $V_{\rm th}(\Phi)$ varies rather slowly.
Thus, $V_{\rm th}$ can be well approximated by a linear
term in the region of interest and one obtains
\begin{eqnarray} 
 V''(\Phi_{\rm crit}) & = &0 \;, \nonumber\\
 T_{\rm crit} & = & \left(- {V'(\Phi_{\rm crit})\over 
 v'_{\rm th}(\Phi_{\rm crit})}\right)^{1/4} \;.
\label{Tcr}
\end{eqnarray}
Note that $V'(\Phi_{\rm crit})$ is the maximal slope of the supergravity 
potential.

\vspace{5mm}

It is important to realize that the moduli are not in thermal equilibrium. 
On the contrary, the moduli interaction rate 
$\Gamma_{\Phi} \sim T^3/M_{\rm P}^2$ is much smaller than the Hubble parameter 
$H \sim T^2/M_{\rm P}$. Hence, the moduli never reach thermal equilibrium. 
In particular, they do  not attain a thermal mass\footnote{This is one of the 
main differences between  our results and  those  of 
Ref.~\cite{Binetruy:1986ss}.}. Rather, they  behave as classical backgrounds.
This is analogous to the behaviour of the gravitational metric in a
thermal bath where the thermal expectation value of the energy momentum tensor
drives the cosmological expansion. On the other hand, gauge 
and/or matter fields are in thermal equilibrium and contribute to the effective
potential through loops. In this way, the gauge plasma exerts a force on the 
classical backgrounds and drives the moduli.

Let us summarize the conditions under which Eq.~(\ref{max}) for the critical
temperature in the radiation dominated phase applies:
\renewcommand{\labelenumi}{{\bf(\roman{enumi})}}
\begin{enumerate}
\item The modulus $\Phi$ is related to some coupling $g$. 
\item The barrier separating a local minimum in $\Phi$
from the `run--away' minimum is related to SUSY breaking. 
\item Gauge and matter fields which couple with strength $g$
are in thermal equilibrium. 
\end{enumerate}
\renewcommand{\labelenumi}{arabic{enumi}}
A few comments are in order. First, $g$ is not necessarily the gauge coupling.
For instance, the Yukawa couplings of twisted states are  functions of the 
$T$--moduli, $Y\sim e^{-\alpha T}$. Then the diagram Fig.~1(c) generates
an effective potential for these moduli, as long as the matter fields are 
in thermal equilibrium. 
Second, if $g$ is a gauge coupling, $\Phi$ is not necessarily the dilaton. 
It can, for instance, be a volume modulus, as it happens in D--brane models.
Third, Eq.~(\ref{max}) holds for positive, negative or zero
cosmological constant, as long as condition (ii) is satisfied.

\clearpage
\noindent
{\large\bf Critical temperature in the KKLT scenario}\\

Background fluxes induce a potential which can stabilize the dilaton and the 
complex structure moduli at appropriate values \cite{Giddings:2001yu}. 
However, at least one K\"ahler modulus, corresponding to the overall volume, 
is not affected by the fluxes and remains undetermined. The volume
can be stabilized by non--perturbative contributions to the superpotential 
such as D--brane instantons or gaugino condensation \cite{Kachru:2003aw}.
The result is a supersymmetric AdS vacuum. 

In order to lift the negative cosmological constant of the AdS vacuum
to a small positive value, extra contributions are added to the scalar 
potential. These can be due to the presence of $\overline{\rm D3}$--branes, 
as in the original KKLT model or SUSY breaking 
D--terms \cite{Burgess:2003ic}.  
A common feature of these modifications is that the new vacuum is 
separated from the `run--away' vacuum \cite{Dine:1985he} by a barrier whose
height is given by the SUSY breaking scale (cf.~Fig.~2).

In D--brane models, the gauge and matter fields of the standard model and the
hidden sector can live on D3 or D7 branes \cite{Camara:2003ku}.
At high temperature, thermalized gauge and matter fields of unbroken 
gauge groups on D7--branes will contribute to the scalar potential of the 
volume modulus. Depending on the D--brane model, the thermalized fields can 
belong to the standard model or to additional
gauge groups for which no gaugino condensation occurs.
This will lead to the existence of a maximal temperature 
beyond which the 
volume modulus is destabilized. In the following we calculate this
critical temperature.

The superpotential and the K\"ahler potential for the volume modulus 
$\rho=\sigma+i\alpha$ are given by \cite{Kachru:2003aw} 
\begin{eqnarray}
 W & = & W_0 + A\, e^{-\,a\,\rho}\;,\nonumber \\
 K & = & -3\,\ln ( 2 \sigma )\;.
\end{eqnarray}
Here $W_0$ is a constant induced by the fluxes, and 
$W-W_0$ represents a  non--perturbative contribution to the superpotential 
due to D--brane instantons or gaugino condensation. In the latter case
the exponent is given by the $\beta$-function of the corresponding gauge
group, $4\pi a = 3/(2\beta)$. The effective 4D Yang--Mills gauge coupling on 
the D7--branes is related to $\sigma$ as
\begin{equation}
\sigma
\ = \ 
{4\pi \over g^2} \;.
\end{equation}
We can always choose the constant $W_0$ to be real and negative. 
$\alpha = {\rm Im}(\rho)$ is then stabilized at a value where
the remaining potential for $\sigma={\rm Re}(\rho)$ reads 
\begin{equation}
 V_0 \ =\ \frac{a\,A\,e^{-a\sigma}}{2\sigma^2}
 \left[\frac{1}{3}\,a\,A\,\sigma \,e^{-a\,\sigma}+W_0+A\,e^{-a\,\sigma}\right]
 \;,
\end{equation}
with $A$ real and positive. $V_0$ has an AdS minimum.
This potential is amended by a supersymmetry breaking term,
\begin{equation}
 V\ =\ V_0 + {D_n \over \sigma^n} \;.
\end{equation}
Here $n=3$ corresponds to the KKLT potential \cite{Kachru:2003aw},
which can be realized as a Fayet-Iliopoulos D--term term
\cite{Burgess:2003ic}, whereas $n=2$ occurs for the explicitly SUSY
breaking contribution of an $\overline{\rm D3}$--brane \cite{Kachru:2003sx}.

\begin{figure}[h]
 \centerline{\CenterObject{\includegraphics[scale=1]{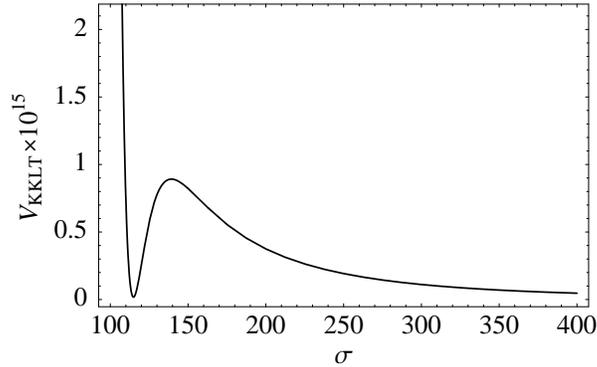}}}
 \caption{ The KKLT potential.}
 \label{fig:VKKLT}
\end{figure}

For certain choices of the parameters, one can obtain dS vacua with a
small cosmological constant. An example is $W_0=-10^{-4}$, $A=1$, $a=0.1$, 
$D_3 = 3\times 10^{-9}$ ($D_2 = 2.6 \times 10^{-11}$); for $n=3$ this is
the KKLT potential. Both cases, $n=3$ and $n=2$ are shown in 
Fig.~\ref{fig:VKKLT}. Numerically, they are almost indistinguishable. 
The local minimum is at $\sigma_{\rm min} \simeq 115$, corresponding to 
the gauge coupling $g=0.3$.

\vspace{5mm}

To calculate the critical temperature, we need the first and the second 
derivative of the potential. For $\sigma \gg 1$ and $a\sigma \gg 1$, 
$V'$ and $V''$ are well approximated by
\begin{eqnarray}
V' &\simeq& -{a^3 A^2 \over 3 \sigma} e^{-2 a \sigma}
-{a^2 A W_0 \over 2 \sigma^2} e^{- a \sigma}\;, \\
V'' &\simeq& {2a^4 A^2 \over 3 \sigma} e^{-2 a \sigma}
+{a^3 A W_0 \over 2 \sigma^2} e^{- a \sigma} \;.
\end{eqnarray}
Note that for the derivatives $D_n/\sigma^n$ is negligible.
Setting $V''=0$, which defines $\sigma_{\rm cr}$, one finds the maximal slope 
of the potential:
\begin{equation}
 V_{\rm max}' 
 \ \simeq \ 
 {a^3 A^2 \over 3 \sigma_{\rm cr}} 
 e^{-2 a \sigma_{\rm cr}}\;.
\label{vp}
\end{equation}

It is straightforward to relate the maximal slope to the scale of
supersymmetry breaking. Since the cosmological constant is negligibly small,
the gravitino mass is given by
\begin{eqnarray}
 m_{3/2}^2 &=& e^K\ |W|^2 \Bigl\vert_{\rm min}\ 
 =\  - {1\over 3}\ V_0(\sigma_{\rm min}) \nonumber\\
 &\simeq& {a^2 A^2 \over 18 \sigma_{\rm min}} e^{-2 a \sigma_{\rm min}}\;.
\label{m32}
\end{eqnarray}
For the KKLT parameters the gravitino mass is very large, 
$m_{3/2} \sim 10^{10}\ {\rm GeV}$.
Note, that in the case of  explicit SUSY breaking, $m_{3/2}$ is not
the physical gravitino mass but just a parameter which controls the scale 
of SUSY breaking. Due to steepness of the potential, $\sigma_{\rm min}$ 
and $\sigma_{\rm cr}$ are very close to each other. Hence, combining 
(\ref{vp}) and (\ref{m32}), one obtains
\begin{equation}
 V_{\rm max}'\ \simeq\ 6 a\,  m_{3/2}^2 \;.
\end{equation}

From Eqs.~(\ref{vth}) and (\ref{Tcr}) one now obtains for the critical 
temperature:
\begin{equation}\label{tcrit}
T_{\rm crit}\ \simeq\ c\, \sqrt{m_{3/2}}\;, 
\end{equation}
with $c\simeq \vert 6a / v_{\rm th}'(\sigma_{\rm cr})\vert^{1/4}
={\cal O}(1)$. Note that this result holds both for 
$\delta V \propto 1/\sigma^3$ and $\delta V \propto 1/\sigma^2$. 
Numerically, for $m_{3/2} \sim 100$ GeV, the maximal temperature is 
\begin{equation}
 T_{\rm crit}\ \sim \ 10^{10}~{\rm GeV} \;. 
\end{equation}

In the case of gaugino condensation one has $4\pi a = 3/(2\beta)$. Using
$1/g^2 = 4\pi \sigma$, Eq.~(\ref{tcrit}) reads explicitly
\begin{equation}
 T_{\rm crit}
 \ \sim\ \sqrt{m_{3/2}} \left({2\over B}\right)^{1/4} 
 \left({3\over \beta}\right)^{1/4} \left({1\over g^2}\right)^{3/8}\;,
\end{equation}
where $B = (1/T^4) \partial F/\partial g(\Phi_{\rm crit})$
\cite{Buchmuller:2004xr}. 
It is instructive to compare this result with racetrack models, where
the K\"ahler potential is also logarithmic and the superpotential is
a sum of two exponential terms. In this case the critical temperature
is given by \cite{Buchmuller:2004xr}
\begin{equation}
 T_{\rm crit}\ \sim\ \sqrt{m_{3/2}} \left({2\over B}\right)^{1/4} 
 \left({3\over \beta}\right)^{3/4} \left({1\over g^2}\right)^{7/8}\;.
\end{equation}
Clearly, both expressions are very similar. The different powers of
$1/\beta$ and $1/g^2$ reflect the differences of the superpotential and
K\"ahler potential in the two cases. For typical values, 
e.g. $1/g^2 \simeq 2$ and $\beta = 0.1$,
the critical temperature in racetrack models is larger by about a factor 
five.\\

\noindent
{\large\bf Discussion}\\

The maximal temperature derived above places a bound on the 
temperature in the radiation dominated phase of the early 
universe. In particular, it bounds the reheating temperature,
\begin{equation}
 T_{\rm reheat}
 \ <\ 
 T_{\rm crit} \;.
\label{reh}
\end{equation}
In addition, it places a bound on the maximal temperature in the preheating
epoch \cite{kt},
\begin{equation}
 T_{\rm preheat}
 \ =\ 
 (T_{\rm reheat}^2 H_{\rm inf} M_{\rm P})^{1/4} 
 \ < \ 
 T_{\rm crit} \;,
\end{equation}
unless the inflaton coupling to $\Phi$ is strong enough
to overcome the destabilizing thermal effects.
This bound is usually more severe than (\ref{reh}),
yet it is also less universal.

We note that these bounds on the temperature of the early universe,
unlike the gravitino  bound, cannot be circumvented by late entropy
production or other cosmological mechanisms. They are also
independent of the `overshoot problem' \cite{Brustein:1992nk}, that during the
cosmological evolution a modulus with a steep potential may not settle
in a shallow minimum, but rather roll over the barrier to infinity. 
This problem can be
solved by tuning the initial conditions or by implementing a mechanism
to slow down the modulus (cf. \cite{Brustein:2004jp} and references
therein). On the contrary, the constraint $T < T_{\rm crit}$ is unavoidable
since there is no local minimum for $T > T_{\rm crit}$
and the modulus inevitably runs away to infinity.

While finalizing this work we received a related paper by Kallosh and Linde
\cite{Kallosh:2004yh} which addresses moduli destabilization during inflation
in KKLT models. The authors also present a model where, at special points
in parameter space, the size of the barrier separating the physical vacuum
from the unphysical one is unrelated to the gravitino mass. However, for
generic parameters, the barrier is related to the scale of supersymmetry
breaking and the analysis of the present paper applies.\\

{\bf Acknowledgements.} 
We would like to thank R.~Brustein for correspondence, and J.~Louis and
H.-P.~Nilles for discussions.
One of us (M.R.) would like to thank the Aspen Center for Physics for support.
This work was partially supported by the EU 6th Framework Program
MRTN-CT-2004-503369 ``Quest for Unification'' and MRTN-CT-2004-005104
``ForcesUniverse''.


\begin{thebibliography}{99}

\bibitem{Kachru:2003aw}
S.~Kachru, R.~Kallosh, A.~Linde and S.~P.~Trivedi,
Phys.\ Rev.\ D {\bf 68} (2003) 046005.

\bibitem{Giddings:2001yu}
S.~B.~Giddings, S.~Kachru and J.~Polchinski,
Phys.\ Rev.\ D {\bf 66} (2002) 106006.

\bibitem{Derendinger:1985kk}
J.~P.~Derendinger, L.~E.~Ib\'a\~nez and H.~P.~Nilles,
Phys.\ Lett.\ B {\bf 155} (1985) 65;
M.~Dine, R.~Rohm, N.~Seiberg and E.~Witten,
Phys.\ Lett.\ B {\bf 156} (1985) 55.

\bibitem{Buchmuller:2003is}
W.~Buchm\"uller, K.~Hamaguchi and M.~Ratz,
Phys.\ Lett.\ B {\bf 574} (2003) 156.

\bibitem{Buchmuller:2004xr}
W.~Buchm\"uller, K.~Hamaguchi, O.~Lebedev and M.~Ratz,
Nucl.\ Phys.\ B {\bf 699} (2004) 292.

\bibitem{Dine:1985he}
M.~Dine and N.~Seiberg,
Phys.\ Lett.\ B {\bf 162} (1985) 299.



\bibitem{Navarro:2004mm}
I.~Navarro and J.~Santiago,
JCAP {\bf 0409} (2004) 005.




\bibitem{Kapusta:1989tk}
J.~I.~Kapusta,
``Finite Temperature Field Theory,''
Cambridge, 1989.




\bibitem{Binetruy:1986ss}
P.~Binetruy and M.~K.~Gaillard,
Phys.\ Rev.\ D {\bf 34} (1986) 3069;
K.~A.~Olive and M.~A.~Srednicki,
Phys.\ Lett.\ B {\bf 148} (1984) 437.


\bibitem{Burgess:2003ic}
C.~P.~Burgess, R.~Kallosh and F.~Quevedo,
JHEP {\bf 0310} (2003) 056.

\bibitem{Camara:2003ku}
P.~G.~Camara, L.~E.~Ib\'a\~nez and A.~M.~Uranga,
Nucl.\ Phys.\ B {\bf 689} (2004) 195;
%
hep-th/0408036;
%
%
M.~Gra\~na, T.~W.~Grimm, H.~Jockers and J.~Louis,
Nucl.\ Phys.\ B {\bf 690} (2004) 21;
%
%
F.~Marchesano and G.~Shiu,
hep-th/0408059;
%
hep-th/0409132;
%
%
H.~Jockers and J.~Louis,
hep-th/0409098;
%
%
D.~L\"ust, S.~Reffert and S.~Stieberger,
hep-th/0410074.

\bibitem{Kachru:2003sx}
S.~Kachru, R.~Kallosh, A.~Linde, J.~Maldacena, L.~McAllister and S.~P.~Trivedi,
JCAP {\bf 0310} (2003) 013.


\bibitem{kt}
E.~Kolb, M.~Turner, The Early Universe, Addison-Wesley, Redwood City, CA 1990.

\bibitem{Brustein:1992nk}
R.~Brustein and P.~J.~Steinhardt,
Phys.\ Lett.\ B {\bf 302} (1993) 196.

\bibitem{Brustein:2004jp}
R.~Brustein, S.~P.~de Alwis and P.~Martens,
hep-th/0408160;
N.~Kaloper, J.~Rahmfeld and L.~Sorbo,
hep-th/0409226.




\bibitem{Kallosh:2004yh}
R.~Kallosh and A.~Linde,
hep-th/0411011.


\end{thebibliography}
\end{document}